\documentclass[twocolumn]{revtex4-1}
\usepackage{bm}
\usepackage{amssymb}
\usepackage{scalerel}
\usepackage{epstopdf}
\usepackage{mathbbol}
\usepackage{amsmath}
\usepackage{graphics,graphicx,epsfig,ulem}

\begin{document}


\title{Mirrorless optical parametric oscillator inside an all-optical waveguide}%

\author{Sushree S. Sahoo}
\email{sushree.ss@niser.ac.in}

\author{Snigdha S. Pati}

\author{Ashok K. Mohapatra}
\email{a.mohapatra@niser.ac.in}
\affiliation{School of Physical Sciences, National Institute of Science Education and Research Bhubaneswar, HBNI, Jatni-752050, India.}


\begin{abstract}
\textbf{Mirrorless optical parametric oscillator (MOPO) is a consequence of intrinsic feedback provided by the nonlinearity in a medium due to the interaction of a pair of strong counter-propagating fields. 
As the name suggests, the device doesn't require a cavity for lasing other than the nonlinear medium. Here, we report the first demonstration of MOPO under the effect of an all-optical waveguide. 
The efficient four-wave mixing process due to counter-propagating pump and control fields interacting with a multilevel atomic system facilitates the generation of mirrorless Stokes and anti-Stokes 
fields counter-propagating to each other.  The maximum generated laser power could rise up to mW with pump conversion efficiency more than 30\%. Furthermore, the cross-phase modulation due 
to the strong Gaussian beams create all-optical waveguides for the generated fields and hence induces different spatial modes in the Stokes as well as the anti-Stokes fields. With suitable experimental parameters, 
we could generate correlated Gaussian mode or Laguerre-Gaussian mode for both the generated fields.}
\end{abstract}

\maketitle

MOPO has attracted a lot attention ever since the phenomenon was first theoretically predicted by S.E. Harris ~\cite{Harris:1966}. It was experimentally demonstrated for the case of parametric down 
conversion in a quasi-phase-matched non-linear $\chi^{(2)}$ medium ~\cite{Canalias:2007aa} and was followed by extensive studies on the coherence properties and conversion efficiency of the 
phenomenon ~\cite{Stromqvist:2011aa, Gustav:2012, Charlotte:2017}. For the case of $\chi^{(3)}$ medium, two counter-propagating strong driving fields result in the spontaneous generation of biphotons 
i.e. Stokes and anti-Stokes field from noise. The photon pairs being generated in opposite directions establish a distributed feedback mechanism assisted by efficient four-wave mixing (FWM). The FWM-based MOPO has been experimentally achieved using electromagnetically induced transparency (EIT)~\cite{Braje:2004aa} and further has been used to generate narrow-band biphotons~\cite{Du:08,Du:2008aa} in cold atomic ensemble. The works by Balic \textit{et} \textit{al.} \cite{Balic:2005aa} and Pavel \cite{Kolchin:2007aa} describe the theory of the counter-propagating biphoton generation showing a good agreement with the experimental results. A recent study in cold atomic ensemble includes the transition of photon correlation properties from the biphoton quantum regime to MOPO regime~\cite{Mei:2017aa}. 

MOPO in thermal atomic vapor has also been a subject of intensive study. Earlier it was used to investigate optical instabilities and self-oscillation by atomic-vapor degenerate FWM ~\cite{Silberberg:1982aa,PInard:1986,Gaeta:1987aa, Khitrova:1988aa}.  Along with the experimental demonstration of MOPO in thermal atomic vapor for the case of non-degenerate FWM~\cite{Zibrov:1999aa}, there has been many theoretical studies on the generated photon pairs~\cite{Lukin:1998aa,Fleischhauer:2000aa,Lukin:1999aa,Jiang:2004aa,Ooi:2007aa}. The phenomenon has also been reported for Raman process in hot atomic vapor~\cite{Zhang:2014aa}.

On the other hand, the nonlinear interaction of a strong beam with medium results in a spatially varying refractive index and hence leads to the formation of an all-optical waveguide. This causes the modification of the spatial profile of a weak probe beam while propagating through the medium. There have been reports on induced focusing ~\cite{Agrawal:1990aa}, spatial-soliton induced waveguides~\cite{DeLaFuente:91,Swartzlander:1992aa}, transverse localization~\cite{Cheng:2005aa,Andre:2005aa,Vengalattore:2005aa} and EIT-induced waveguides
~\cite{Moseley:1996aa,Kapoor:2000aa,Shpaisman:2005aa}. Optically written waveguides in thermal Rubidium vapor cell have been achieved using Gaussian pump beam~\cite{Andersen:2001aa} as well as doughnut-shaped pump beam~\cite{Truscott:1999aa,Vudyasetu:2009aa}. 

In this work, we demonstrate MOPO in thermal vapor with counter-propagating strong driving fields called as pump and control beams. The Gaussian profiles of the strong fields result in a nonlinear refractive index-induced wave-guide for the generated fields and hence lead to the generation of wave-guided spatial modes. Although there have been substantial works on the study of gain and threshold of the FWM-based MOPO in atomic vapor, the other aspect like the spatial modes of the generated biphotons have not been investigated before. We report the first-ever experimental demonstration of the spatial correlation between the photon pairs generated via the MOPO process with a very high pump conversion efficiency. 


\begin{figure}[t]
\begin{center}
\includegraphics[width=80mm,scale=1]{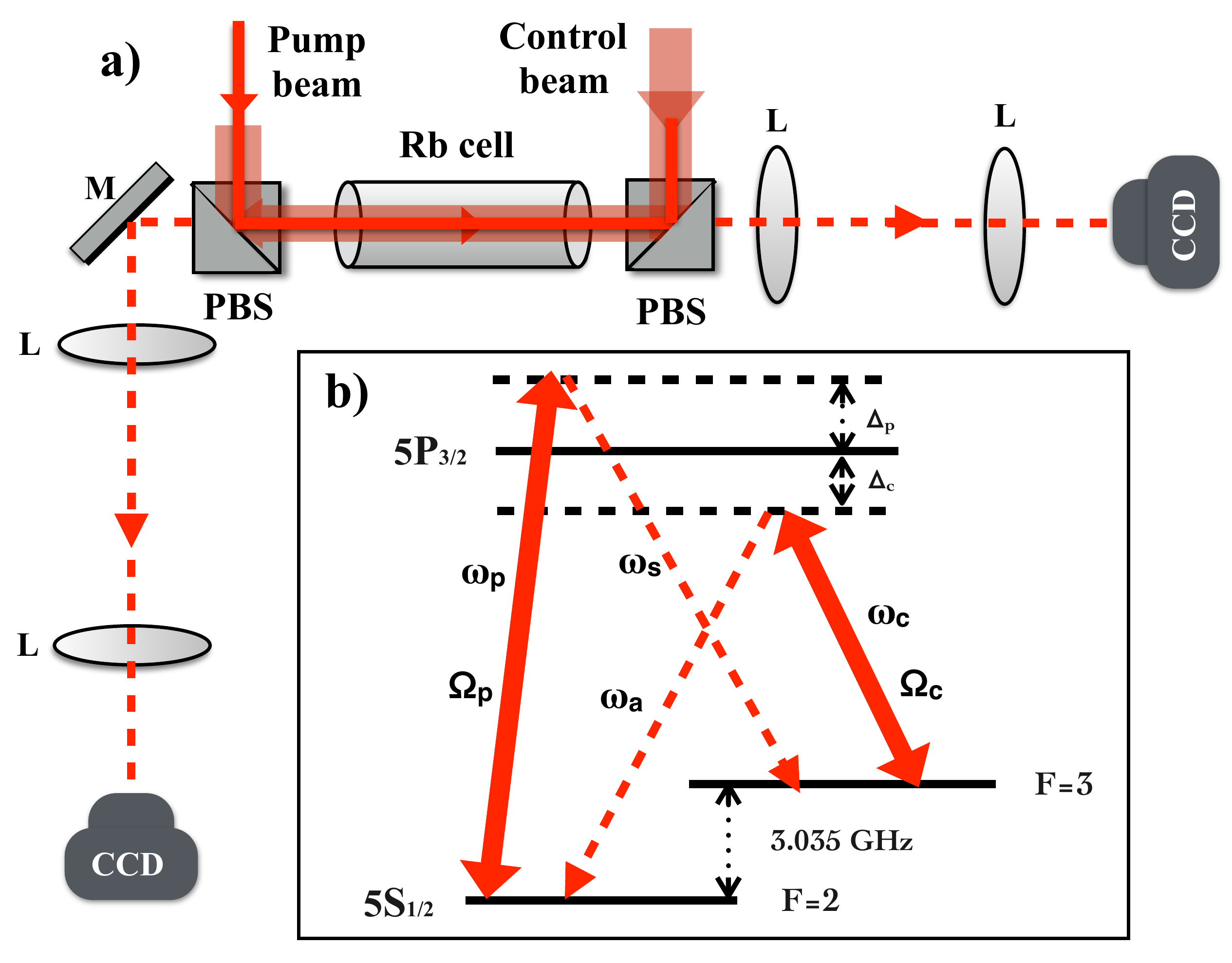}
\caption{a) Experimental setup for MOPO in thermal rubidium vapor, PBS: polarising beam splitter, M: mirror, L: Lens. b) Energy level diagram of $^{85}$Rb showing the coupling of pump and control laser fields 
(solid lines) and the generated Stokes and anti-Stokes fields (dashed lines).}
\label{fig.energy}
\end{center}
\end{figure}

The schematic of the experimental setup and the relevant energy level diagram for the system are shown in Fig. ~\ref{fig.energy}. A pump laser field with Rabi frequency $\Omega_p$ couples the transition $^5$S$_{1/2}$, $F=2$ $\rightarrow$ 5P$_{3/2}$ of $^{85}$Rb with it's optical frequency $\omega_p$ stabilized at about 1.2 GHz blue-detuned to the transition. Similarly, the control laser field with Rabi frequency $\Omega_c$ and optical frequency $\omega_c$ is detuned to the red of the transition $^5$S$_{1/2}$, $F=3$ $\rightarrow$ 5P$_{3/2}$ by 800 MHz. The Stokes and the anti-Stokes fields with respective frequencies $\omega_s$ and $\omega_{a}$ are spontaneously generated and amplified due to non-degenerate FWM process in the medium while obeying the energy conservation i.e. $\omega_s+\omega_{a}=\omega_p+\omega_c$. The efficient FWM process assisted by the hyperfine ground state coherence of the system provides the intrinsic feedback for the generated fields. Both the strong fields are linearly polarised in the same direction and the generated fields are measured in the orthogonal polarisation direction. The $\frac{1}{e}$ radii of the pump and the control beams used in the experiment are 200$\mu$m and 1mm respectively.

As a first observation of the signal, we used a Fabri-Perot cavity to measure the frequency of the generated Stokes beam with respect to the pump beam. A typical signal from a cavity is presented 
in the supplementary material~\cite{supp} to confirm the separation between them to be approximately equal to the hyperfine splitting, $\nu_{\text{\tiny{HF}}}$ given by 3.035 GHz. The pump threshold is found to depend on 
the control Rabi frequency as well as the laser detunings~\cite{Mei:2017aa} and in our experimental condition, the threshold Rabi frequency is found to be $\simeq$30 MHz when the control Rabi frequency is 140MHz. A study on the variation of the Stokes power with the input strong fields is discussed in the supplementary material~\cite{supp}. The maximum Stokes power is found to be $\simeq$300 $\mu$W with 30$\%$ pump conversion efficiency for the above-mentioned beam waists of the pump and control beams. However, using a bigger input pump beam i.e. with beam waist $\simeq$ $0.5$ mm, we could generate the Stokes power up to an mW with the similar conversion efficiency.

\begin{figure}[t]
\begin{center}
\includegraphics[width=85mm,scale=1]{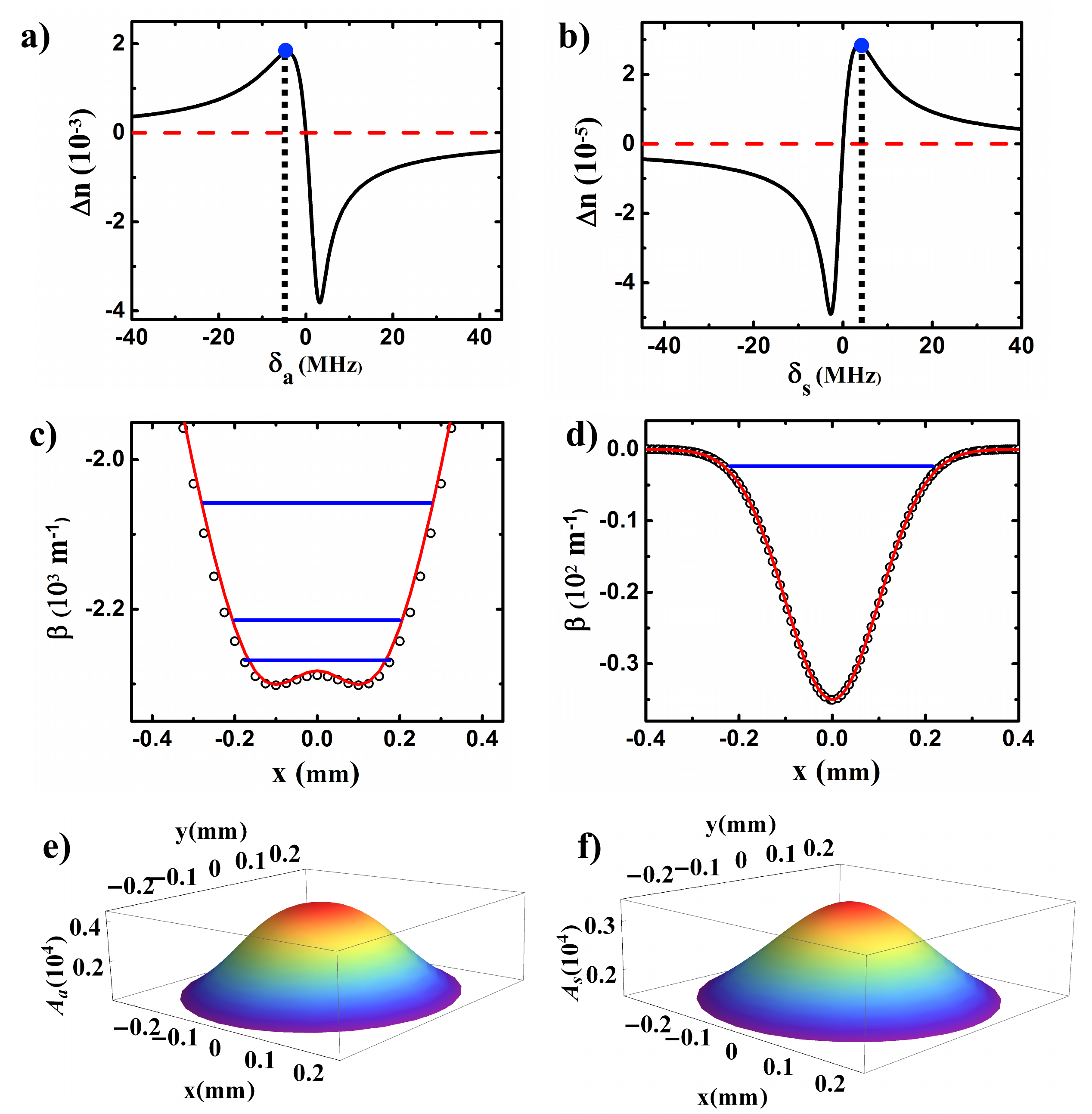}
\caption{Nonlinear refractive index ($\Delta$n) as a function of the two-photon detuning, a) $\delta_a$ for anti-Stokes and b) $\delta_s$ for Stokes fields. 
Transverse profile of the nonlinear refractive indices experienced by the generated fields due to Gaussian intensity profiles of the pump and control fields were calculated at specified points in the respective figures. 
Nearly Gaussian potentials experienced by c) anti-Stokes and d) Stokes fields. The black circles represent the potentials evaluated from the theoretical model and the red solid line signifies the functional fitting of the form, $a0(1-e^{-x^2/(0.68)^2}+0.065e^{-x^2/(0.142)^2})$ and $s0(1-e^{-x^2/(0.142)^2})$ for the anti-Stokes and Stokes beams respectively. The horizontal red solid lines on the potentials denote the corresponding eigenvalus (theoretically calculated). The normalized ground-state eigenfunction of the waveguide for the case of e) anti-Stokes field and f) Stokes fields. }
\label{fig.ref}
\end{center}
\end{figure}

To understand the non-linearity experienced by the generated fields in the system, we introduce an external probe beam with the same frequency as the generated Stokes beam and study it's transmission 
through the medium under the similar experimental condition as for the case of MOPO generation. The pump and the control beams are slightly misaligned such that the FWM process is inefficient 
with a phase mismatch and the lone effect of the cross phase modulation can be studied. As theoretically studied and reported in references~\cite{Verma15,Islam17}, the presence of the pump and the control fields induces a nonlinear gain as well as the non-linear refractive index in the probe beam due to interaction 
with an N-type system. We experimentally measure the probe transmission as a function of the two-photon detuning and observe a nonlinear gain in the probe field~\cite{supp}. In order to study the effect of the nonlinear dispersion, we analyze the spatial profile of the probe beam while varying the two-photon detuning and it is found to be focused or defocused depending on the value of the nonlinear refractive index. We further perform the theoretical calculation for the given N-system to estimate the dispersion of the Stokes as well as the anti-Stokes field with the respective pump and control laser parameters. Taking into account the
laser detunings as well as the Rabi frequencies of the control and the pump fields, the Stokes beam experiences a gain whereas the anti-Stokes beam undergoes absorption due to the nonlinear interaction. The details of the theoretical study can be found in the supplementary material~\cite{supp}.
 
Fig.~\ref{fig.ref}(a) and (b) represent the theoretically calculated nonlinear refractive indices ($\Delta$n) for the Stokes and the anti-Stokes beams as functions of their respective laser detunings for near-threshold MOPO. Owing to the fact that the control
Rabi frequency is larger as compared to the pump Rabi frequency, the value of $\Delta$n for the anti-Stokes beam is always greater than that for the Stokes beam. Here, the respective two-photon detunings for the Stokes and the anti-Stokes beams are given by the expressions, $\delta_s=\omega_s-\omega_p-\Delta_{\text{\tiny{LS}}}+\nu_{\text{\tiny{HF}}}$ and $\delta_a=\omega_a-\omega_c+\Delta_{\text{\tiny{LS}}}-\nu_{\text{\tiny{HF}}}$ with $\Delta_{\text{\tiny{LS}}}$ being the light shift due to the strong fields. The energy conservation associated with the FWM process i.e. $\omega_s+\omega_{a}=\omega_p+\omega_c$ ensures that $\delta_s+\delta_a=0$. Therefore, if the anti-Stokes beam is generated at $\delta_a$, the Stokes beam is generated at $\delta_s=-\delta_a$ while both of them experiencing a positive nonlinear refractive index as specified by the blue points in Fig.~\ref{fig.ref}(a) and (b) and in turn lead to MOPO under the effect of the all-optical wave guide. We theoretically calculate the spatially varying refractive index potentials as experienced by the generated beams and found that they are approximately Gaussian in nature. The control beam with a larger width forms a wider Gaussian waveguide than the pump beam and the corresponding Gaussian potentials along one of the transverse direction are presented in Fig.~\ref{fig.ref}(c) and (d). 

The spatial modes of the generated fields under the effect of the all-optical waveguides can be studied using their propagation equations through the nonlinear medium. As discussed in the method section, the nonlinear gain of the generated fields can be 
ignored to study the spatial modes of the generated fields inside the non-linear waveguide and the corresponding wave equations for the generated beams can be written as,
$$i\frac{\partial A_{s,a}}{\partial z}=-\frac{1}{2k_{s,a}}\nabla_{T}^2A_{s,a}+V_{s,a}(r)A_{s,a}.$$
Where $A_s$ and $A_a$ are the amplitudes of the Stokes and anti-Stokes beams respectively, $r= \sqrt{x^2+y^2}$ is the radial co-ordinate, $z$ is the propagation direction and $V_{s,a}(r)=-\Delta n_{s,a}(r)$ 
are the near-Gaussian potentials as experienced by the Stokes and anti-Stokes fields respectively.

The propagation equation resembles the Schrodinger's equation with a two-dimensional Gaussian potential. We numerically solve the equation to find the eigenmodes of the potentials for both the generated 
beams. The theoretically estimated eigenmodes are depicted in Fig.~\ref{fig.ref}(c) and (d) for the potentials along one transverse component. As discussed in the method section, for efficient MOPO under the effect of an all-optical waveguide, 
the eigenmodes of the waveguide should meet the condition of phase-matching inside the medium as,
  $$\beta_{a}-\beta_s+\Delta k=0.$$

Apart from the phase-matching condition, a large overlapping integral is necessary for the efficiency of the process. To understand this, we consider the case of near-threshold MOPO. As depicted in Fig.~\ref{fig.ref}(c) and (d), the waveguide for the Stokes beam has one spatial eigenmode whereas the anti-Stokes beam can support many eigenmodes. For both the waveguides, we plot the ground-state eigenfunction and found them to be similar as shown in Fig.~\ref{fig.ref}(e) and (f), which ensures a large overlapping integral. Similarly, it can be shown that the overlapping integral is also larger for the next higher-order mode. But, it is observed that the eigenfunctions differ more and more with higher-order modes and hence leads to a reduction in MOPO process. As understood theoretically and verified experimentally, the generation of the spatial modes, especially the higher-order modes are critical with regard to the beam sizes of the strong driving fields. 

\begin{figure}[t]
\begin{center}
\includegraphics[width=85mm,scale=1]{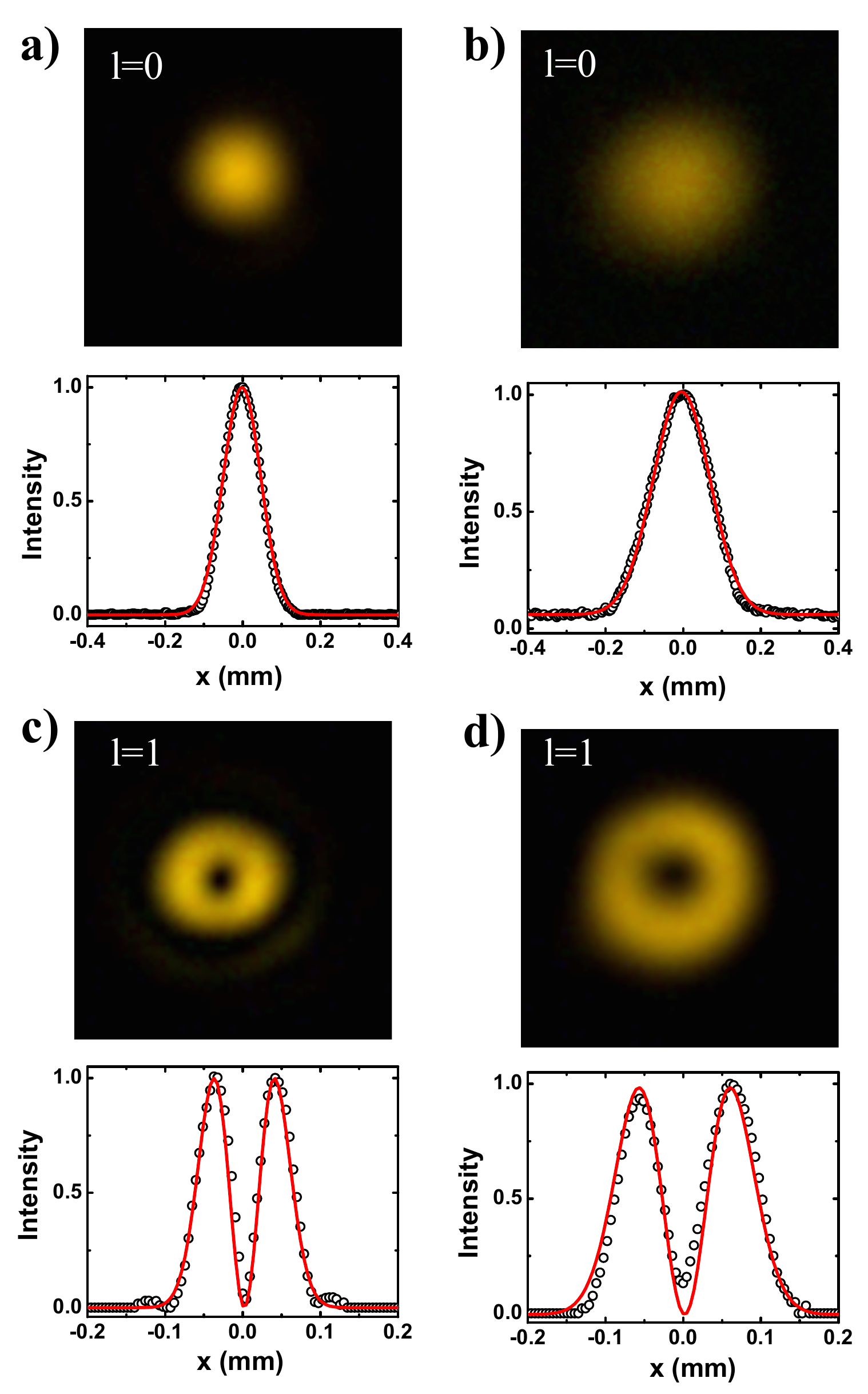}
\caption{The symmetric gaussian mode and the corresponding transverse intensity patterns in the generated a) Stokes and b) anti-Stokes fields. The first-order Laguerre-Gaussian (LG01) mode and the corresponding transverse intensity patterns in the generated c) Stokes and d) anti-Stokes fields.}
\label{fig.sm}
\end{center}
\end{figure}

To analyze the spatial profiles of the generated beams, the beams are imaged to the CCD camera at the exit face of the vapor cell in both directions. The control Rabi frequency is kept fixed at 140MHz and the transverse profiles of the Stokes and anti-Stokes beams are recorded by varying the pump Rabi frequency. We observe that with a very low pump Rabi frequency i.e. $\simeq$ 30MHz, biphotons with Gaussian spatial mode are generated and eventually with an increase in pump Rabi frequency, we observe the generation of Laguerre-Gaussian (LG) mode. In Fig.~\ref{fig.sm} (a) and (b), we show the CCD images of the Gaussian mode with the corresponding intensity profile along the transverse co-ordinate x for Stokes and anti-Stokes beams respectively. The intensity profile is normalized to the peak value of intensity and is fitted with a Gaussian distribution function. The different beam sizes for both the generated beams could be due to the fact that they are imaged at different positions of the cell i.e. at the back face and front face of the cell for Stokes and anti-Stokes beam respectively. Similarly, Fig.~\ref{fig.sm} (c) and (d) refers to the CCD images as well as the transverse intensity patterns of Laguerre-Gaussian (LG) mode for both the generated beams which was achieved with pump Rabi frequency at 60MHz. The corresponding intensity patterns match well with that of a first order Laguerre-Gaussian mode(LG01). The Stokes and the anti-Stokes beams are found to be spatially correlated in addition to their temporal correlation. Furthermore, the observation of the far field images of the generated fields confirms their propagation in the same spatial modes.

To investigate the temporal modes of the MOPO, a part of the Stokes beam is fed to a fast photodetector and is analyzed using an RF spectrum analyzer.  For higher pump powers i.e. $\simeq$ 70-120 MHz, we observe a frequency comb with a spacing of the order of a few KHz to MHz, which is understood to be the result of the Zeeman coherence induced forward four-wave mixing~\cite{Sahoo17} due to the generation of Stokes beams with two different frequencies as explained in the supplementary material~\cite{supp}. The narrow lines of the frequency comb suggests that the frequencies of the temporal modes of the Stokes beam are highly correlated with each other and hence their individual frequency is highly correlated with the pump frequency. The phase-matching condition ensures that a single temporal mode corresponds to a single spatial mode, which is experimentally verified by the fact that no frequency comb structures is observed in the regimes of single spatial modes. Hence, we generate single spatio-temporal modes of the MOPO inside an all-optical waveguide.

\begin{figure}[t]
\begin{center}
\includegraphics[width=85mm,scale=1]{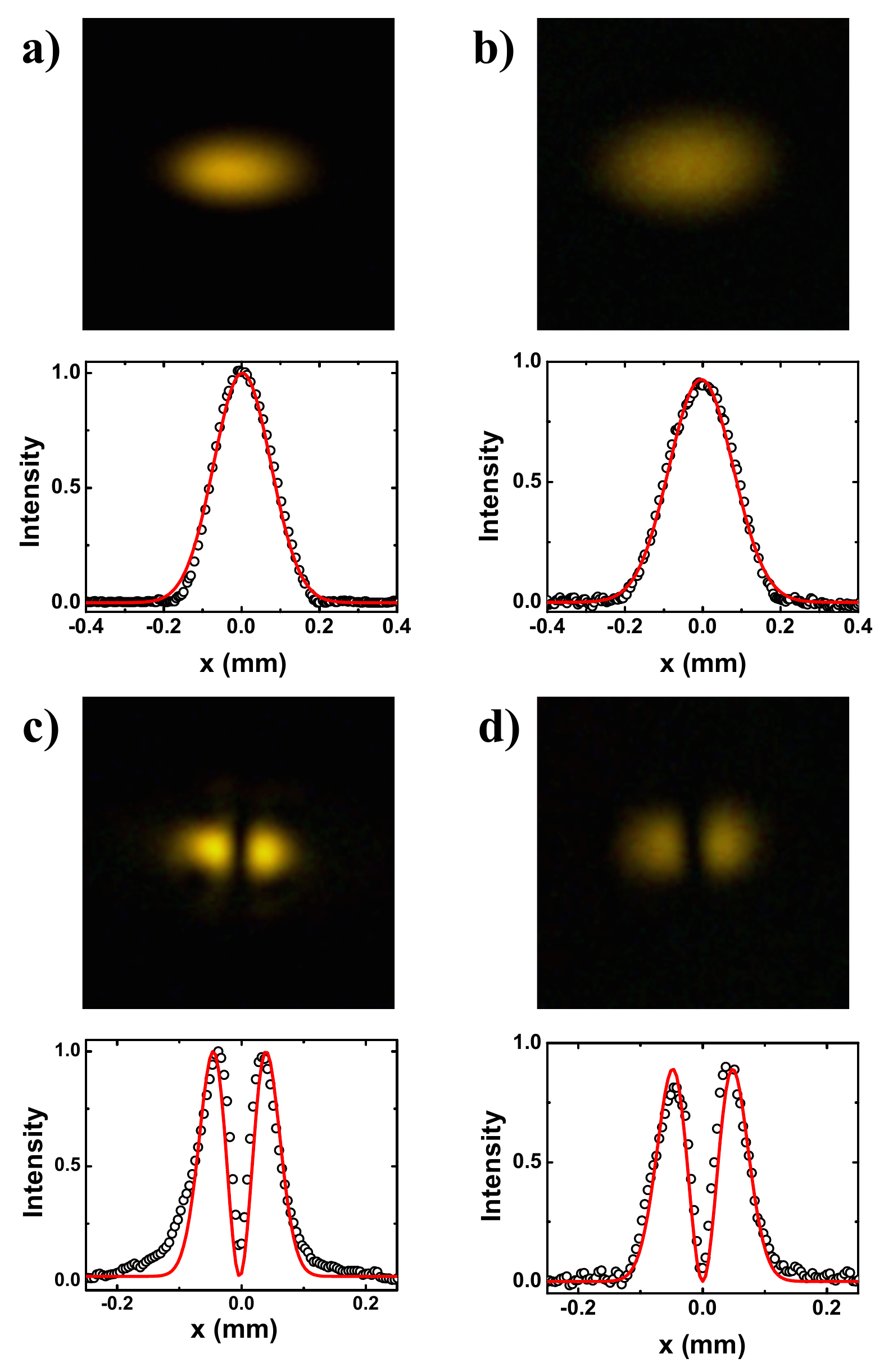}
\caption{The asymmetric gaussian mode and the corresponding transverse intensity patterns in the generated a) Stokes and b) anti-Stokes fields. The first-order Hermite-Gaussian (HG01) mode and the corresponding transverse intensity patterns in the generated c) Stokes and d) anti-Stokes fields.}
\label{fig.am}
\end{center}
\end{figure}

As a further extension of the experiment, the symmetry of the waveguide along the transverse directions is broken by using an elliptical pump beam. A pair of anamorphic prisms are used to make the pump profile elliptic with ellipticity $\simeq$1.5. The same experiment is repeated to observe different spatial modes of the generated beams i.e. by varying the pump Rabi frequency. The input asymmetric pump beam induces an asymmetric Gaussian potential for the generated fields and hence leads to the generation of asymmetric Gaussian mode near threshold as shown in Fig.~\ref{fig.am}(a) and (b). We have fitted the intensity patterns along one transverse direction with Gaussian distribution function as a confirmation of the spatial mode. In this case, as the cylindrical symmetry of the waveguide is broken, the next higher order eigenmode is the Hermite-Gauss mode. As expected, we observed the same mode in the experiment with an increase of the pump Rabi frequency. The corresponding experimental data is presented in Fig.~\ref{fig.am}(c) and (d). The spatial patterns generated in the experiment are quite repeatable provided it is under the suitable parameter regime. 


In this work, we have studied the generation of all-optically guided MOPO which is found to be due to the combined effect of FWM and XPM in a system with efficient nonlinear processes. We report the generation of spatially as well as temporally correlated biphoton pairs with a very high pump conversion efficiency. Our further study includes optimization of the system to get stable higher modes of the all-optical wave guide. Additionally, we have observed a bistability phenomenon in the threshold of the single spatio-temporal modes of the generated beams. So our future interest is to study and understand the origin of this bistability. Moreover, we plan to work on the mirrorless laser to increase its efficiency by feedback of the generated beam into the nonlinear medium. It also would be interesting to study the effect of the disorder 
introduced in the nonlinear refractive index experienced by the generated fields by using the random intensity pattern of the pump or control fields similar to the study of random lasers~\cite{wier08,schi12}.

\section{Methods}

\subsection{Experimental methods}
The layout of the experimental setup is presented in the Fig. \ref{fig.energy} (a). Here the pump and control laser beams are derived from two different external cavity diode lasers tuned to $780$nm. The pump beam is passed through a single mode fiber (SMF) for mode cleaning to get a nearly Gaussian beam profile. The control beam from the other laser is aligned in the opposite direction to the pump beam using a polarising beam splitter (PBS) so that both the beams counter-propagate each other inside a 5 cm long vapor cell housed in two layers of magnetic shield and are linearly polarised in the same direction. With the counter-propagating pump and control beams, the phase-matching condition results in generated beams being counter-propagating to each other. The beam which is generated along the direction of the pump beam is called as the Stokes beam and the one generated along the control beam is called the anti-Stokes beam. The Stokes beam is then split into three parts using two non-polarising beam splitters (NPBS). One part is fed to a Fabri-Perot cavity, the second part is fed to a spectrum analyzer and the last part is fed to a CCD camera. Here we have used 4-f imaging technique to image the generated beam at the exit-face of the vapor cell to the camera. The same detection procedure is followed for the anti-Stokes beam as well in the opposite direction. The temperature of the vapor cell is kept between $110^0-120^0$C using a controlled heater which corresponds to a number density of approximately $10^{13}/$cm$^3$. The pump and the control powers are varied by using a combination of half-wave plate and PBS to achieve different parameter regimes. 

\subsection{Wave equations for the generated fields} 

The wave equation of the Stokes field propagating along the z-direction with the radial transverse co-ordinate $r$ is given as,
\begin{eqnarray*}
\nabla^2E_s(r)+\frac{\omega_s^2}{c^2}\left(1+\chi^{(1)}\right)E_s(r)=\frac{-\omega_s^2}{\epsilon_0c^2}P^{(NL)}(r) 
\end{eqnarray*}
$P^{(NL)}(r)=3\epsilon_0 \chi_{\text{\tiny{XPM}}}^{(eff)}(r)E_s+3\epsilon_0\chi_{\text{\tiny{FWM}}}^{(3)}E_p E_c E_{a}$ which is 
the non-linear polarization of the medium oscillating with frequency $\omega_s$ of the Stokes field and  $\chi^{(1)}$ is the linear susceptibility, which is related to the linear refractive index as, $n_s=\sqrt{\left(1+\chi^{(1)}\right)}$. Replacing the electric field amplitudes in the propagation equation by, $E_i=A_ie^{in_ik_iz}$, i=p,c,s,a with $k_i$ being the wave-vector and $n_i$ being the linear refractive index associated with the corresponding field. With the slowly varying amplitude approximation, the wave equation 
now can be reduced to

\begin{widetext}
\begin{eqnarray}
2ik_s\frac{\partial A_s}{\partial z}+\nabla_T^2A_s+2i\alpha_sE_s +3k_s^2Re\left( \chi_{\text{\tiny{XPM}}}^{(eff)}(r)\right)A_s+3k_s^2\chi_{\text{\tiny{FWM}}}^{(3)}A_p A_c A_{a}^*e^{i\Delta kz}=0 
\end{eqnarray}
\end{widetext}
 
Here,
$\Delta k=n_pk_p-n_ck_c-n_{s}k_s+n_{a}k_{a}$ is the linear phase-mismatch in the system. $\chi_{\text{\tiny{XPM}}}^{(eff)}$ is the effective susceptibility of the probe beam due to the N-system, where the real part contributes to the nonlinear refractive index experienced by the probe due to the presence of the control and pump fields and hence is responsible for the wave guiding of the generated fields. $\alpha_s=\frac{k_s}{2}Im\left(\chi_s^{(1)}+3\chi_{\text{\tiny{XPM}}}^{(eff)}\right)$ accounts for the gain of the Stokes beam in the medium and $\chi_{\text{\tiny{FWM}}}^{(3)}$ is the nonlinear susceptibility due to the non-degenerate four wave mixing process. 
The transverse profiles of the generated fields in the lowest order mode of the all-optical waveguides would be much smaller than the transverse size of the control and pump fields. In this case, the pump and the
control fields can be considered as plane waves and hence the gain due to the four-wave mixing and the nonlinear absorption process would result in a uniform gain/loss in the transverse direction of the generated fields. 
Hence, it would be a good approximation to use that the spatial profiles of the generated fields are only decided by 
the all optical waveguide while considering at least the lowest order mode. In order to determine the spatial mode of the generated field, it is safe to neglect the four wave mixing term in the above equation as
 $i\frac{\partial A_s}{\partial z}=\beta_sA_s$  with $\beta_s=-\frac{\nabla_T^2}{2k_s}+V(r)$. $V(r)=-\frac{3}{2n_s}Re(\chi_{\text{\tiny{XPM}}}^{(eff)}(r))$ is the potential experienced by the Stokes field with $\chi_{\text{\tiny{XPM}}}^{(eff)}$ being the value of the effective susceptibility at the peak of the potential. If $V(r)$
is independent of $z$, then $\beta_s$ is constant. A similar equation can be derived for the anti-Stokes field with the corresponding potential. If the generated fields are excited in one of the eigen modes of the waveguide, then the solution of the
spatial modes would be $A_s=A_{s0}e^{i\beta_sz}$ and $A_a=A_{a0}e^{-i\beta_az}$ for the counter-propagating Stokes and anti-Stoke fields respectively. Replacing the spatial mode solution in equation (1), we get
\begin{eqnarray*}
ik_s\frac{\partial A_{s0}}{\partial z}+i\alpha_sA_{s0}=-\kappa_sA_{a0}^*e^{i\left(\beta_a-\beta_s+\Delta k\right)z}\\
-ik_a\frac{\partial A_{a0}}{\partial z}+i\alpha_aA_{a0}=-\kappa_aA_{s0}^*e^{i\left(\beta_a-\beta_s+\Delta k\right)z}
\end{eqnarray*}
where $\kappa_s=\frac{3}{2}k_s^2\chi_{\text{\tiny{FWM}}}^{(3)}A_pA_c$ and $\kappa_a=\frac{3}{2}k_a^2\chi_{\text{\tiny{FWM}}}^{(3)}A_pA_c$. The phase matching condition required for efficient generation of the field due to four-wave mixing inside the waveguide to be $\beta_a-\beta_s+\Delta k=0$.

\textbf{Acknowledgment.} The authors gratefully acknowledge useful discussions with Dr. T. N. Dey, D. Kara, A. Bhowmick and T. Firdoshi. 
This work was financially supported by the Department of Atomic Energy, Govt. of India. 

\textbf{Author contributions.} A.K.M. conceived the concept of the mirrorless laser in the presence of all-optical wave-guide. S.S.S. and A.K.M. contributed to the theoretical model. S.S.S. and S.S.P. performed the 
experiment and analysed the data. All the authors contributed in preparing the manuscript.

\textbf{Data Availability Statement.} Correspondence and requests for data that supports the observation presented in this paper should be addressed to A.K.M. (email: a.mohapatra@niser.ac.in).

\textbf{Competing Interests:} The authors declare that they have no financial competing interests.

\vspace{50px}

\textbf{\large{Supplementary Information}}

\begin{figure}[htbp]
\begin{center}
\includegraphics[width=80mm,scale=1]{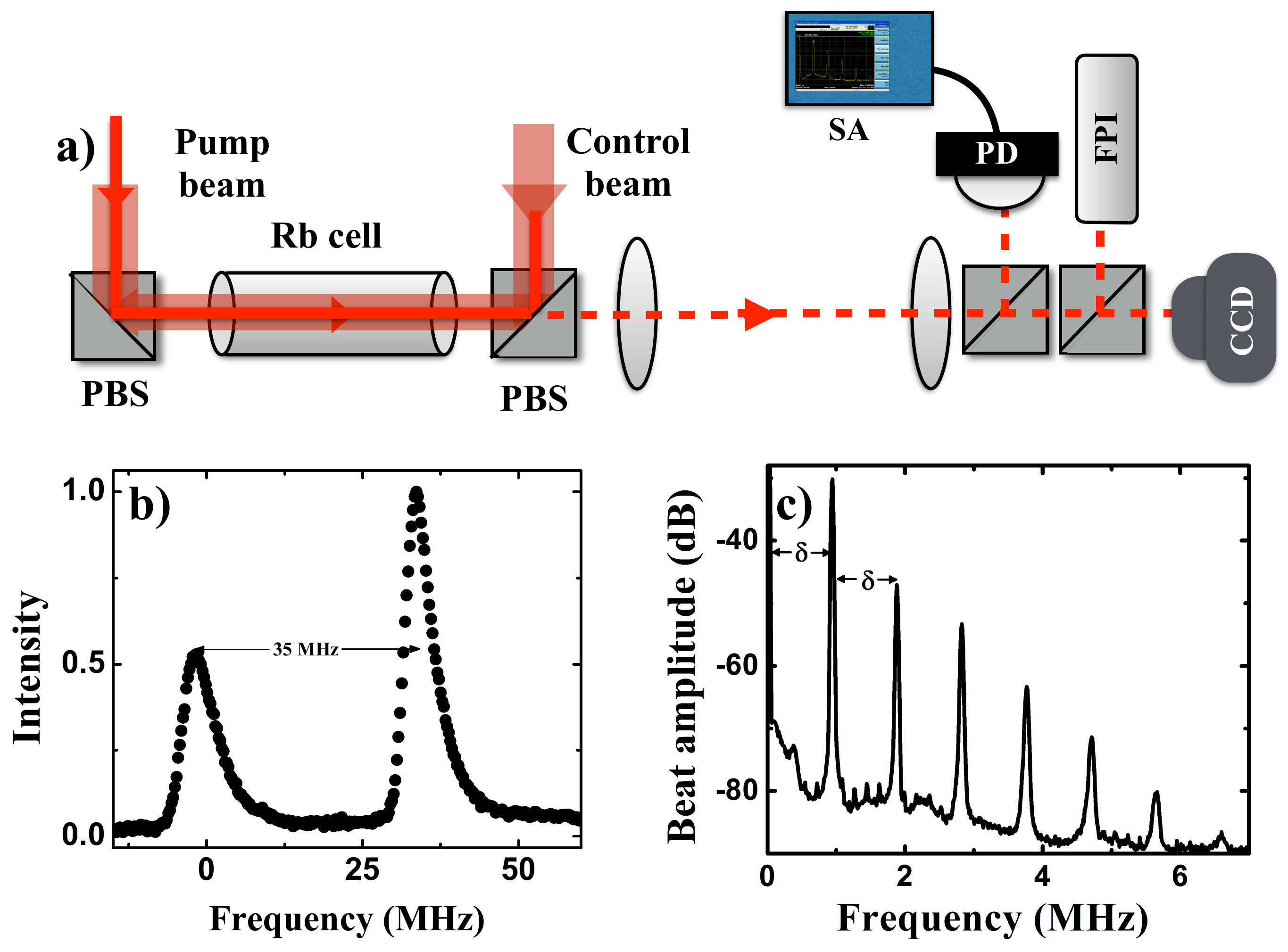}
\caption{ Schematic of the experimental setup. b) Cavity signal showing the frequency difference of about 35 MHz between the pump and the Stokes beam using a Fabri-Perot cavity of free spectral range of 1 GHz. c) Spectrum analyzer signal showing the frequency comb under multi-spatial-mode regime. }
\label{fig.expt}
\end{center}
\end{figure}

\textbf{Supplementary Methods: Experimental details}

As mentioned in the main text, in order to confirm the generation of the Stokes and anti-Stokes beam in the MOPO process, we send a part of the generated light into a Fabri-Perot cavity. Fig.~\ref{fig.expt} (a) shows the experimental setup and Fig.~\ref{fig.expt} (b) shows a typical cavity signal for near-threshold MOPO. The small peak corresponds to a small leakage pump beam and the larger peak is due to the generated beam. With the free spectral range of the FPI being 1 GHz, the frequency difference of $35$ MHz between the peaks confirms the generation of the Stokes beam being frequency-separated from the pump beam by the hyperfine difference of $3.035$ GHz. The broad cavity signal is a contribution from the losses in the cavity rather than the laser-phase noise.

Furthermore, we investigate the multi-spatial mode regime of the MOPO, which is observed for pump Rabi frequency in the range, $\simeq$ 70-120 MHz. We send a part of the Stokes beam into a fast photodetector, which is connected to a spectrum analyzer. In this regime, there is generation of at least two different temporal modes of the Stokes beam. These beams being the lower modes of the waveguide have beam waist of the order of $50$ $\mu$m and with the experimentally observed beam power of $\simeq 150$ $\mu$ W, the corresponding Rabi frequency is as large as $100$ MHz. Therefore, the beams undergo efficient forward four-wave mixing due to Zeeman degenerate two-level system~\cite{Sahoo17} and lead to the generation of equi-spaced frequency comb. Fig.~\ref{fig.expt} (c) depicts the spectrum analyzer signal showing the frequency comb with fundamental frequency $\delta_c$. The value of $\delta_c$ can be tuned from hundreds of KHz to few MHz and strongly depends on the alignment of the strong driving beams as well as their Rabi frequencies and laser detunings. The same frequency comb structure is observed for the anti-Stokes beam simultaneously. Under certain parameter condition, we have observed multiple simultaneous frequency comb structures, which could be due to the generation of three or more Stokes beams.

\begin{figure}[htbp]
\begin{center}
\includegraphics[width=80mm,scale=1]{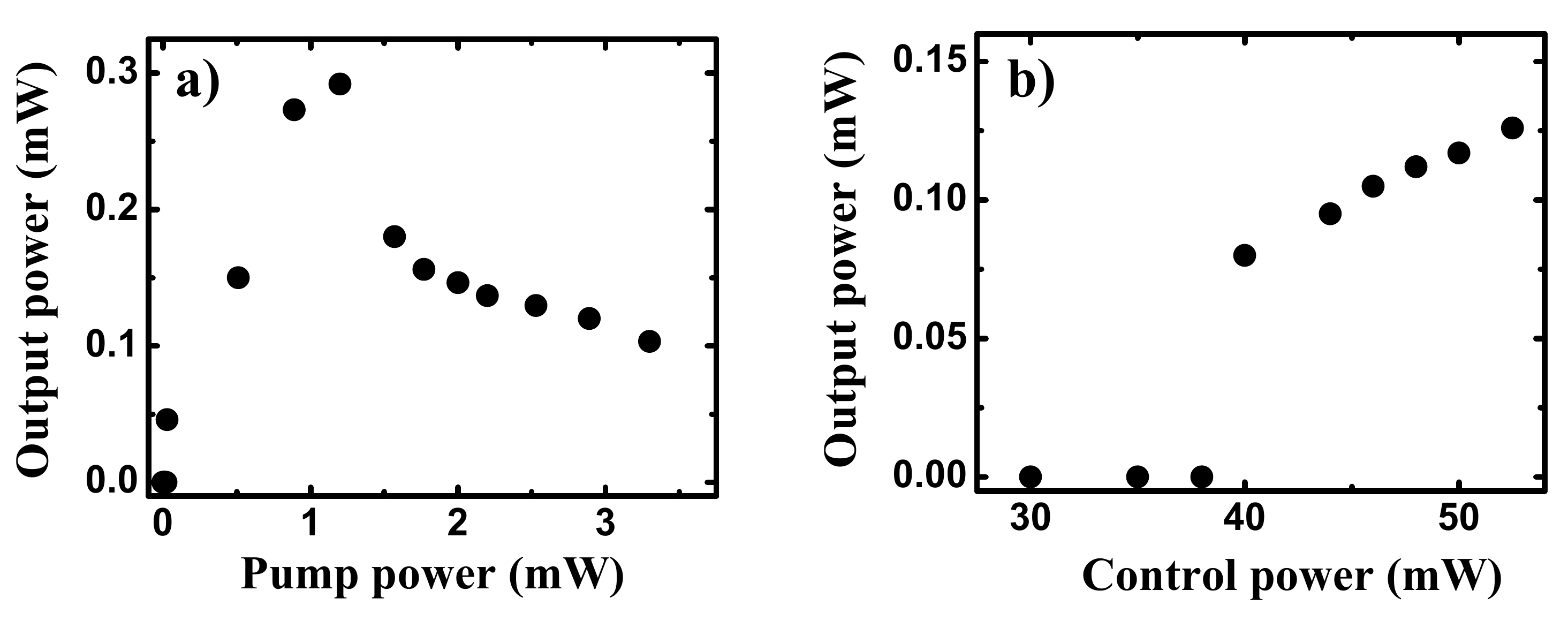}
\caption{ Output generated Stokes power a) as a function of pump power with control power fixed at 55mW and b) as a function of control power with pump power fixed at 1.3mW.}
\label{fig.thr}
\end{center}
\end{figure}

In order to study the threshold characteristics of the MOPO system, we measured the generated Stokes power as a function of the input pump power for control power fixed at $55$ mW as well as by varying control power while pump power was fixed at 1.3mW. The beam sizes of the pump and control beams used are 200$ \mu$m and 1mm respectively corresponding to the Rabi frequencies of 140 MHz for 55 mW control power and 90 MHz for 1.3 mW of pump power. The corresponding experimental data is presented in Fig.~\ref{fig.thr}. From the experiment, the pump threshold value was found to be $0.27$mW. We could get the maximum probe power of $\simeq$ 300$\mu$W with input pump power as 0.9mW, where the corresponding spatial mode was first-order Laguerre-Gaussian for the generated light. However using a larger pump beam size i.e. $0.5$ mm, we could get upto a mW of power in the Stokes field with more than 30$\%$ of pump-conversion efficiency.

\begin{figure}[htbp]
\begin{center}
\includegraphics[width=80mm,scale=1]{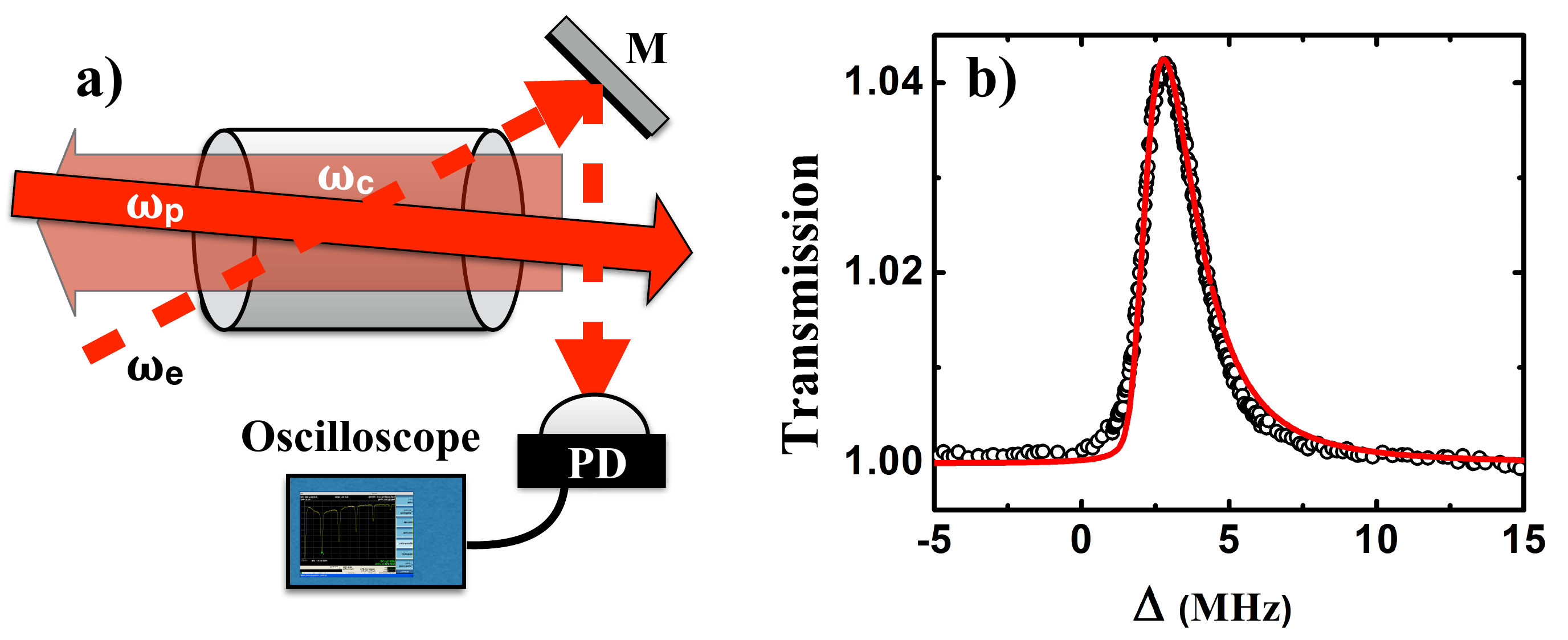}
\caption{a) Schematic of the experimental setup for probe propagation in N-type system. b) Matching of experimentally measured and theoretically evaluated transmission of the external probe. The black circles represent the experimental data and the red solid line represent the data from the theoretical model.}
\label{fig.trans}
\end{center}
\end{figure}

The generated counter-propagating fields by the MOPO process undergo cross-phase modulation due to the presence of the strong beams. In order to investigate it experimentally, we set up an experiment by putting an external probe into the system of counter-propagating pump and control beams. The schematic of the experimental setup is presented in Fig.~\ref{fig.trans} (a). The probe beam is derived from the same laser as that of the pump beam and the frequency separation of 3.035 GHz is achieved by double-pass AOM configuration of 1.5 GHz. The strong beams are slightly misaligned in order to avoid the MOPO process. In this case, we send the transmitted probe beam into a photo-detector, which is connected to an oscilloscope. We measured the probe absorption or gain in the medium by scanning the probe frequency using the AOM.

\begin{figure}[htbp]
\begin{center}
\includegraphics[width=80mm,scale=1]{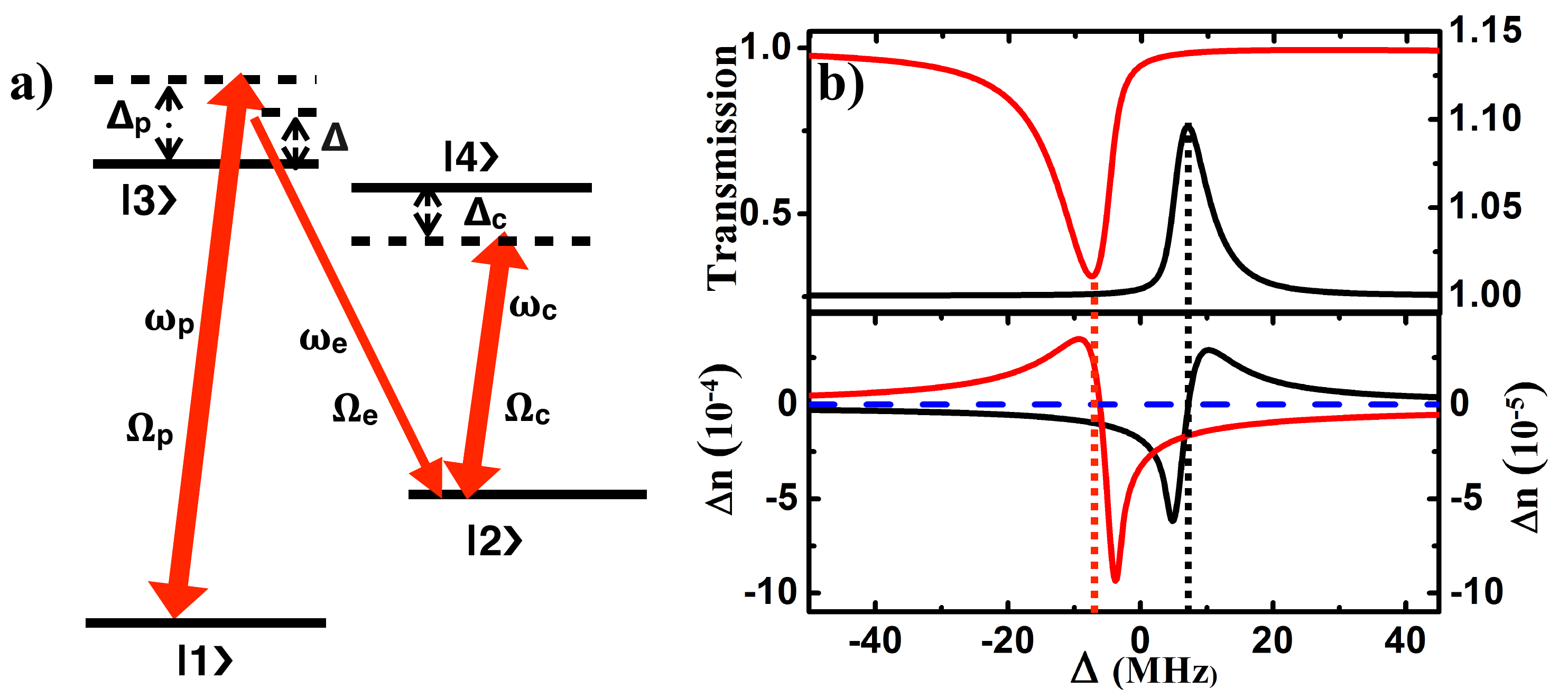}
\caption{a) Schematic of the energy level diagram of the N-type system. b) Transmission and nonlinear refractive index($\Delta$ n) of the Stokes and anti-Stokes beams. The red solid line corresponds to the anti-Stokes beam and the black solid line corresponds to the Stokes beam. The scaling of the Transmission and $\Delta n$ for the anti-Stokes beams are presented in the left-axis of the graph whereas that of the Stokes beam are presented in the right axis. The blue dotted line corresponds to the line of null refractive index.}
\label{fig.theory}
\end{center}
\end{figure}

\textbf{Theoretical model}:

The schematic of the four-level N-type system with two ground states and two excited states is presented in Fig.~\ref{fig.theory}(a). The system is acted upon by three input light fields i.e. pump, control and an external probe field with Rabi frequencies $\Omega_p$, $\Omega_c$ and $\Omega_e$ respectively. The external probe field couples the transition as that of one of the generated beams of the MOPO process. The pump and control beams are counter-propagating and the direction of the probe field is chosen according to whether it signifies the Stokes field or the anti-Stokes field. The single-photon detunings for the pump, control and probe fields are given by $\Delta_p$, $\Delta_c$ and $\Delta_e$, whereas $\Delta$ is defined by, $\Delta_e-\Delta_p$. The system has been analyzed both theoretically and experimentally to study the narrow absorptive resonance with gain as well as the dispersion of the probe beam~\cite{Bason09, Islam17}. The same system has further been used for optical steering, cloning and splitting of probe beam~\cite{Verma15}.

Here, we present an alternate model by considering a system of three levels with pump and probe fields while the contribution of the control field is implemented by modification in the populations and coherences. With semi-classical approach and rotating-wave approximation, the Hamiltonian of the three level system with pump and the probe field is given by,

\[
\Tilde{H}=-\frac{\hslash}{2}
\begin{pmatrix}
0&  {\Omega_{e}}& 0\\
\Omega_{e}^{*}& -2\Delta_p& \Omega_{p}\\
0&   \Omega_{p}^{*}& -2\Delta
\end{pmatrix}
\]

The time evolution of the system is described by the master equation:
\begin{equation}
i\hslash
\frac{d\rho}{dt}=[H,\rho]+i\hslash L_{D}
\end{equation} 
Here $\rho$ stands for the density matrix operator and $L_{D}$ stands for the Linblad operator, which describes all the decay and dephasing rates in the medium. The system is associated with a decay rate $\Gamma$ of $6$MHz from the excited state to the ground states and a dephasing rate $\gamma_c$ of $0.1$ MHz between the ground levels. 

The master equation results in six optical Bloch equations of the density matrix elements which are solved in steady state by considering the equilibrium populations of the ground states in the presence of the 
strong pump and control fields. The population distribution in the hyperfine states depends on the optical pumping rates due to both the strong fields provided that all the population decay rates from the 
excited states to the ground states are same. Hence, the value of $(\rho_{11}^{(eq)}-\rho_{22}^{(eq)})$ is calculated by using $\frac{\rho_{11}^{(eq)}}{\rho_{22}^{(eq)}}=\frac{\Omega_c^2}{\Omega_p^2}(\frac{2\Omega_p^2+4\Delta_p^2+\Gamma^2}{2\Omega_c^2+4\Delta_c^2+\Gamma^2})$, as derived by considering the two-level optical pumping rates of the pump and the control fields.

  Considering the Doppler broadening in the medium, the laser detunings $\Delta_p$, $\Delta_c$ and $\Delta_e$ are modified as $\Delta_p-kv$, $\Delta_c+kv$ and $\Delta_e-kv$ respectively. With this effect, the nonlinear susceptilibity of the weak probe beam $\chi_{\text{\tiny{XPM}}}^{(eff)}$, evaluated under the steady state condition is doppler-avaraged to get, $\chi_{\text{\tiny{XPM}}}^{(eff)}= \frac{2n|\mu|^2}{\hbar \epsilon_0\Omega_e}\frac{1}{\sqrt{2\pi v_p}}\int_{-\infty}^{\infty}  \rho_{32}(v)e^{-v^2/{2v_p^2}} dv$, where $n$ is the vapor density of rubidium, $\mu$ is the transition dipole moment and  $v_p$ is the most probable speed of the atoms. The real part of the doppler-avaraged susceptibility provides information about the nonlinear gain or absorption in the system while the real part contributes to the nonlinear refractive index as experienced by the generated fields.
Fig.~\ref{fig.trans}(b) shows the matching of the experimentally measured probe gain signal with the theoretical model for near-threshold MOPO, where Rabi frequency of the pump beam is very low. So the model is able to explain the nonlinearity experienced by the Stokes beam beam passing through the medium. As for the case of anti-Stokes beam, the roles of the pump and control beams are interchanged in the model. We have calculated the transmission and nonlinear refractive index of the Stokes and anti-Stokes beams as presented in Fig.~\ref{fig.theory} (b). The theoretical parameters are $\Omega_p=60$ MHz, $\Omega_c=140$MHz, $v_p=190$ m/s and $T=110^0$C.

\end{document}